\documentclass{article}
\usepackage{t1enc}      

\usepackage{amsmath}
\usepackage{amssymb}
\usepackage{mathrsfs}   

\newcommand{\uA}{\underline A \,}
\newcommand{\uB}{\underline B \,}
\newcommand{\uC}{\underline C \,}

\newcommand{\bi}{\bf i} 
\newcommand{\bj}{\bf j} 

\numberwithin{equation}{section}

\begin{document}
\bibliographystyle{unsrt}

\title{A note on the Hamiltonian constraint in canonical GR}

\author{L\'aszl\'o B. Szabados\\ 
        Research Institute for Particle and Nuclear Physics\\
        H-1525 Budapest 114, P. O. Box 49\\ 
        Hungary\\
        e-mail: lbszab@rmki.kfki.hu}
\maketitle

\begin{abstract}
The Hamiltonian constraint of the coupled Einstein--Yang--Mills--Higgs 
system with a cosmological constant is shown to be a pure Poisson 
bracket of a dimensionless functional on the phase space and the 
volume of the three-space. One of its potential consequences, a 
restriction on the eigenstates of the volume operator in a class of 
canonical quantum gravity theories, is also pointed out. 
\end{abstract}


\section{Introduction}
\label{sec-1}

It has been conjectured for a long time that the Chern--Simons 
functional, introduced originally in pure differential geometry 
\cite{CS}, should play some fundamental role in physics (see e.g. 
\cite{BM}). In particular, the conformally invariant functional on 
Riemannian 3-manifolds \cite{C86} can be generalized for initial data 
sets of general relativity: the Chern--Simons functional built from 
an appropriate connection on the pull back to the Cauchy hypersurface 
of the spacetime tangent bundle is a global conformal invariant of 
the initial data set \cite{BSz,Sz00}. The functional can also be 
introduced in the canonical formulation of general relativity too, 
where its conformal invariance implies the vanishing of its Poisson 
bracket with the 3-volume. 

The Chern--Simons functional can be defined on the spinor bundle over 
the Cauchy hypersurface too. In this case it is complex, its real part 
is the previous functional, but its imaginary part is {\em not} a 
conformal invariant. However, this is connected with the vacuum general 
relativity: the Chern--Simons functional, defined in the spinor 
representation, is invariant with respect to infinitesimal conformal 
rescalings on every Cauchy surface precisely when the vacuum Einstein 
equations are satisfied \cite{Sz02}. This made it possible to 
reformulate the Hamiltonian constraint of vacuum GR: this constraint 
is the Poisson bracket of the spinor Chern--Simons functional and 
the 3-volume. 

Recently Soo generalized the above result to include the cosmological 
constant, and he discussed its implications in a quantum version of 
canonical general relativity \cite{S07}, too. 

In the present paper we extend this result further by showing that the 
Hamiltonian constraint even for the coupled 
Einstein--Yang--Mills--Higgs system (with or without the cosmological 
constant) is the Poisson bracket of an appropriate dimensionless 
functional and the volume of the three-space. In a canonical quantum 
theory of gravity this yields a potential restriction on the volume 
eigenstates of the quantum volume operator. 

Our conventions followed here are those of \cite{Sz02}. In particular, 
the three-metric is {\em negative} definite, the spacetime curvature 
is defined by $-{}^4R^a{}_{bcd}X^b:=2\nabla_{[c}\nabla_{d]}X^a$, when 
Einstein's equations take the form ${}^4G_{ab}=-\kappa T_{ab}$ with 
$\kappa:=8\pi G$, and the orientation of the induced volume 3-form of 
the spacelike hypersurface $\Sigma$ with future pointing unit normal 
$t^a$ is defined by $\varepsilon_{abc}:=\varepsilon_{abcd}t^d$.


\section{The Hamiltonian constraint}
\label{sec-2}

For the sake of simplicity the base manifold $\Sigma$ is assumed to be 
closed. The canonical variables are $(q_{ab},\tilde p^{ab})$, $(\phi
^{\bi},\tilde\pi_{\bi})$ and $(A^{\bi}_a{}_{\bj},\tilde E^{a{\bi}}{}
_{\bj})$ in the gravitational, the Higgs and the Yang--Mills sectors, 
respectively. The Hamiltonian constraint of the coupled 
Einstein--Yang--Mills--Higgs system with the cosmological constant 
$\lambda$ is 

\begin{equation}
\tilde{\cal C}:=\tilde{\cal C}_0+\frac{\lambda}{\kappa}\sqrt{\vert q
\vert}+\mu_H\sqrt{\vert q\vert}+\mu_{YM}\sqrt{\vert q\vert}=0, 
\label{eq:2.1}
\end{equation}
where $\tilde{\cal C}_0$ is the Hamiltonian constraint function of the 
vacuum GR and $\mu_H$ and $\mu_{YM}$ are the energy densities of the 
Higgs and Yang--Mills fields, respectively, given explicitly by 

\begin{eqnarray}
\tilde{\cal C}_0\!\!\!\!&:=\!\!\!\!&-\frac{1}{2\kappa}\Bigl(R-
 \frac{4\kappa^2}{\vert q\vert}\bigl(\tilde p^{ab}\tilde p^{cd}q_{ac}
 q_{bd}-\frac{1}{2}[\tilde p^{ab}q_{ab}\bigr]^2\bigr)\Bigr)\sqrt{\vert 
 q\vert}, \label{eq:2.2.a} \\
\mu_H\!\!\!\!&:=\!\!\!\!&\frac{1}{2}G^{\bi\bj}\pi_{\bi}\pi_{\bj}-
 \frac{1}{2}G_{\bi\bj}q^{ab}\bigl({\mathbb D}_a\phi^{\bi}\bigr)\bigl(
 {\mathbb D}_b\phi^{\bj}\bigr)+U\bigl(\phi\bigr), \label{eq:2.2.b} \\
\mu_{YM}\!\!\!\!&:=\!\!\!\!&-\frac{1}{2}q^{ab}\Bigl(E^{\bi}_a{}_{\bj}
 E^{\bj}_b{}_{\bi}+B^{\bi}_a{}_{\bj}B^{\bj}_b{}_{\bi}\Bigr). 
 \label{eq:2.2.c}
\end{eqnarray}
Thus the gravitational canonical variables $(q_{ab},\tilde p^{ab})$ 
are the ADM variables \cite{ADM}, $R$ is the curvature scalar, and 
$\tilde p^{ab}$ is built from the metric and the extrinsic curvature 
$\chi_{ab}$ of $\Sigma$ in the spacetime as $\tilde p^{ab}=-\frac{1}
{2\kappa}(\chi^{ab}-\chi q^{ab})\sqrt{\vert q\vert}$. 
The Higgs field is a multiplet $\{\phi^{\bi}\}$, ${\bi}=1,...,n$, of 
real scalar fields on $\Sigma$, which are transformed among each other 
under the action of an $n$-dimensional representation of some compact 
gauge group. This representation of the gauge group is assumed to be 
a subgroup of $GL(n,{\mathbb R})$ leaving the symmetric positive 
definite metric $G_{\bi\bj}$ fixed. The canonical momentum $\tilde\pi
_{\bi}$ is just $\sqrt{\vert q\vert}$-times $\pi_{\bi}$, where the 
latter is given in terms of the Lagrange variables $(\phi^{\bi},\dot
\phi^{\bi})$ and the lapse $N$ and the shift $N^e$ by $\pi_{\bi}=-
\frac{1}{N}G_{\bi\bj}(\dot\phi^{\bj}-N^e{\mathbb D}_e\phi^{\bj})$. 
${\mathbb D}_e$ is the gauge covariant derivative operator: ${\mathbb 
D}_e\phi^{\bi}:=D_e\phi^{\bi}+A^{\bi}_{e{\bj}}\phi^{\bj}$, and $U
(\phi)$ is some potential, e.g. typically of the form $\frac{1}{2}
m^2G_{\bi\bj}\phi^{\bi}\phi^{\bj}+\frac{1}{4}\nu(G_{\bi\bj}\phi
^{\bi}\phi^{\bj})^2$ with the so-called rest mass $m$ and 
self-interaction parameter $\nu$. 
Finally, the canonical momentum $\tilde E^{a{\bi}}{}_{\bj}$ for the 
Yang--Mills field is $\sqrt{\vert q\vert}$-times the electric field 
strength $E^{a{\bi}}{}_{\bj}$ , and $B^{a{\bi}}{}_{\bj}$ is the 
magnetic field strength. 
The other constraints (namely the momentum constraint of GR and the 
Gauss constraint of the Yang--Mills theory) will not play any role 
in the present paper. 

On the gravitational sector of the phase space we introduce the real 
valued functions 

\begin{equation}
V\bigl[N\bigr]:=\int_\Sigma N\sqrt{\vert q\vert}{\rm d}^3x, 
\hskip 20pt
T\bigl[f\bigr]:=\frac{2}{3}\int_\Sigma f\tilde p^{ab}q_{ab}{\rm 
d}^3x \label{eq:2.3}
\end{equation}
for any real, integrable functions $N$ and $f$ on $\Sigma$. If $D
\subset\Sigma$ is any measurable set and $N$ is its characteristic 
function, then $V[N]$ is the 3-volume of $D$. Essentially $V[N]$ is 
Misner's time function (more precisely, it is $-\frac{1}{3}
\ln{V[1]}$), while $T[f]$ is the smeared version of York's time 
function. Their Poisson bracket is $\{T[f],V[N]\}=V[fN]$. 

Let us fix a spinor structure on $T\Sigma$. Then the intrinsic 
Levi-Civita connection $D_e$ and the extrinsic curvature $\chi_{ab}$ 
determine a connection ${\cal D}_e$ on the spinor bundle, the 
so-called Sen connection, according to ${\cal D}_e\lambda^A:=D_e
\lambda^A-\frac{1}{\sqrt2}\chi_e{}^A{}_B\lambda^B$, where the second 
index of the extrinsic curvature $\chi_{ef}$ has been converted to 
a (symmetric) pair $AB$ of unitary spinor indices. Denoting the 
corresponding connection 1-form and curvature 2-form in some 
normalized dual spin frame $\{\varepsilon^A_{\uA},\varepsilon^{\uA}
_A\}$, ${\uA}=0,1$, by $\Gamma^{\uA}_{e{\uB}}$ and $F^{\uA}{}_{{\uB}
cd}$, respectively, the Chern--Simons functional is defined by 

\begin{equation}
Y\bigl[\Gamma^{\uA}{}_{\uB}\bigr]:=\int_\Sigma\Bigl(F^{\uA}{}_{{\uB}
de}\Gamma^{\uB}_{f{\uA}}+\frac{2}{3}\Gamma^{\uA}_{d{\uB}}\Gamma^{\uB}
_{e{\uC}}\Gamma^{\uC}_{f{\uA}}\Bigr)\frac{1}{3!}\delta^{def}_{abc}.
\label{eq:2.4}
\end{equation}
For its basic properties (in particular the change under the 
transformation of the spinor basis $\{\varepsilon^A_{\uA},
\varepsilon^{\uA}_A\}$ or the conformal rescaling of the spacetime 
metric, the calculation of its functional derivatives with respect 
to the canonical variables as well as a more detailed discussion of 
the geometric background) see \cite{Sz02}. What we need here is the 
result that $Y$ modulo $8\pi^2$ is invariant with respect to the 
change of the spinor basis (and hence ${\rm Re}\,Y$ modulo $8\pi^2$ 
and ${\rm Im}\,Y$ are well defined real valued functions on the ADM 
phase space), and that 

\begin{equation}
\Bigl\{{\rm Re}\,Y,V\bigl[N\bigr]\Bigr\}=0, \hskip 20pt
\Bigl\{{\rm Im}\,Y,V\bigl[N\bigr]\Bigr\}=\kappa^2\int_\Sigma\tilde
{\cal C}_0N{\rm d}^3x; \label{eq:2.5}
\end{equation}
i.e. the Hamiltonian constraint of the vacuum Einstein theory is the 
pure Poisson bracket of the Chern--Simons functional built from the 
Sen connection on the spinor bundle and Misner's time \cite{Sz02}. 

This result can be extended to Einstein--Yang--Mills--Higgs systems. 
In fact, if we define 

\begin{equation}
G:={\rm Im}\,Y+T\bigl[\kappa\lambda\bigr]+\kappa^2\frac{2}{3}\int
_\Sigma\bigl(\mu_H+\mu_{YM}\bigr)\tilde p^{ab}q_{ab}{\rm d}^3x, 
\label{eq:2.6}
\end{equation}
then by the Poisson bracket of the Misner and York times, equation 
(\ref{eq:2.5}) and the definitions it follows that 

\begin{equation}
\Bigl\{G,V\bigl[N\bigr]\Bigr\}=\kappa^2\int_\Sigma\tilde{\cal C}N
{\rm d}^3x=:\kappa^2\,H\bigl[N\bigr], \label{eq:2.7}
\end{equation}
which is a generalization of the previous results of \cite{Sz02} 
for the vacuum, and of Soo \cite{S07} for the cosmological constant 
cases. Thus the geometric content of the Hamiltonian constraint is 
that $G$ must be constant along the flow of the Hamiltonian vector 
field of $V[N]$. $G$, being dimensionless and depending on no 
smearing function, appears to be the `universal generator function' 
by means of which the constraint governing the time evolution of the 
Einstein--Yang--Mills--Higgs system is generated. The lapse 
function $N$ enters the dynamics only through $V[N]$. 

Neither $G$ nor $V[N]$ has weakly vanishing Poisson bracket with the 
constraints, and, in particular, with the Hamiltonian constraint. 
Thus they are {\em not} classical observables on the {\em whole} phase 
space. However, these Poisson brackets could be zero on certain {\em 
subsets} $U$ of the constraint surface (e.g. at the points representing 
Einstein's static universe with a cosmological constant), and in this 
they behave as well defined classical observables on the special 
states represented by the points of $U$.


\section{A restriction on the eigenstates of the quantum volume 
operator}
\label{sec-3}

The result (\ref{eq:2.7}) can be reformulated in Ashtekar's phase 
space, the starting point of most of the recent approaches of 
canonical quantum gravity \cite{Ash}. Indeed, for the vacuum and the 
vacuum with cosmological constant cases this is already given in 
\cite{Sz02} and \cite{S07}, respectively. (In fact, Soo used the even 
more general Barbero--Immirzi variables as the basic canonical 
coordinates.) Thus what remains be done is to express the metric $q
^{ab}$ in $\mu_H$ and $\mu_{YM}$ and the coefficient $\tilde p^{ab}
q_{ab}$ of $\mu_H$ and $\mu_{YM}$ in the generator function $G$ in 
terms of the Ashtekar variables, which is a straightforward 
calculation. 

In the present section we intend to point out a simple consequence 
of (\ref{eq:2.7}) in a class of canonical quantum theories of general 
relativity that are based on Dirac's quantization of constrained 
systems. 
First, it is known that well defined quantum operators for both the 
area of a surface and the volume of a compact domain $D$ in the 
three-space $\Sigma$ can be introduced, and the spin network states are 
eigenstates of them \cite{RoSm1,RoSm2}. As we mentioned in connection 
with (\ref{eq:2.3}), any domain $D$ can be characterized by its own 
characteristic function $N$, and the corresponding volume operator, 
acting as a linear operator on some complex representation space 
${\cal V}$, will be represented in the same way and will be denoted by 
$\widehat{V[N]}$. Next we will have three assumptions: 
1. We assume that the classical generator functional $G$ and the 
Hamiltonian constraint function $H[N]$ have a well defined operator 
form, $\widehat{G}$ and $\widehat{H[N]}$, respectively, acting on 
${\cal V}$. (In fact, what we use in the subsequent discussion is that 
they are well defined on the volume eigenstates.) 
2. Suppose that the classical Poisson bracket relation (\ref{eq:2.7}) 
still holds in operator form: $[\widehat{G},\widehat{V[N]}]={\rm i}
\hbar\,\kappa^2\,\widehat{H[N]}$. (Indeed, as Soo already showed 
\cite{S07}, in the connection representation for an appropriate 
(symmetric) ordering there is an operator form $\widehat{H[N]}$ of 
the Hamiltonian constraint of the vacuum GR which is the commutator 
of the Chern--Simons operator and the three-volume. Hence {\em in 
this representation} our first two assumptions are satisfied.) 
3. Following Dirac, we consider a state $\vert\Psi\rangle$ to be a 
{\em physical state} if and only if it is annihilated by the operator 
form of all constraints (and the space of these states will be 
denoted by ${\cal V}_0$), i.e. in particular $[\widehat{G},\widehat{V
[N]}]\,\vert\Psi\rangle=0$ must hold for any physical state $\vert
\Psi\rangle$. 

Now let us consider an eigenstate $\vert\Psi_v\rangle$ of the volume 
operator $\widehat{V[N]}$ with eigenvalue $v$. Then the action of the 
operator form of (\ref{eq:2.7}) on the eigenstate $\vert\Psi_v
\rangle$ gives 

\begin{equation}
\widehat{V\bigl[N\bigr]}\Bigl(\widehat{G}\vert\Psi_v\rangle\Bigr)=
v\Bigl(\widehat{G}\vert\Psi_v\rangle\Bigr)-{\rm i}\hbar\,\kappa^2\,
\widehat{H[N]}\vert\Psi_v\rangle. \label{eq:3.1}
\end{equation}
Since in general $\widehat{V[N]}$ is not (weakly) commuting with the 
constraint operators, the volume eigenstates are not expected to be 
physical states, and hence the second term on the right hand side of 
(\ref{eq:3.1}) is not vanishing. (The Hamiltonian constraint is not 
required to annihilate the volume eigenstates even if the volume is 
expected to be a quantum observable in the sense of Kucha\v{r} 
\cite{Ku}.) However, on certain subspaces ${\cal U}\subset{\cal V}
_0$ the volume operator could be commuting with the constrains, and 
hence certain volume eigenstates could be physical states as well. 

Thus suppose that $\vert\Psi_v\rangle$ is a physical state too, and 
hence it is annihilated by the Hamiltonian constraint operator. 
Therefore, by (\ref{eq:3.1}) {\em the volume eigenstate $\vert\Psi_v
\rangle$ can be a physical state only if the operator $\widehat{G}$ 
maps the eigenstate $\vert\Psi_v\rangle$ into another eigenstate 
with the same eigenvalue $v$}. (To have a necessary and sufficient 
condition the other constraints also would have to annihilate $\vert
\Psi_v\rangle$.) However, in general, without additional restrictions 
on $\widehat{G}$, the operators $\widehat{G}$ and $\widehat{V[N]}$ 
are {\em not} necessarily simultaneously diagonalizable even on the 
subspace spanned by the eigenstates $\vert\Psi_v\rangle$ with fixed 
$v$; and even if, in addition, every state in this subspace were a 
physical state: 
$\widehat{G}$ may still have a non-trivial Jordan form there. 
This restriction on the structure of the operator $\widehat{G}$ in 
these special states may help finding the operator form of the 
universal generator function $G$ (and, in particular, of the 
Chern--Simons functional) in other (e.g. the loop) representations. 

\medskip

The author is grateful to Chopin Soo, Korn\'el Szlach\'anyi and 
P\'eter Ve\-cser\-ny\'es for the very useful, enlightening 
discussions on the quantum theoretic issues; and to the referees 
for their helpful criticism. This work was partially supported by 
the Hungarian Scientific Research Fund (OTKA) grants T042531 and 
K67790.

\end{document}